\pdfoutput=1

\documentclass[11pt]{article}

\usepackage[preprint]{acl}

\usepackage{times}
\usepackage{latexsym}
\usepackage{comment}

\usepackage[T1]{fontenc}

\usepackage[utf8]{inputenc}

\usepackage{microtype}

\usepackage{inconsolata}

\usepackage{graphicx}

\usepackage{subcaption}
\usepackage{amsmath}
\usepackage{amsfonts}
\usepackage{tabularray}
\UseTblrLibrary{booktabs}
\SetTblrInner[booktabs]{abovesep=0pt, belowsep=0pt, rowsep=0.5pt}
\SetTblrInner[booktabs]{cells = {cmd=\small}}

\NewTableCommand\seprule{\specialrule{\lightrulewidth,gray8}{2.5pt}{2.5pt}}
\NewTableCommand\uniquerule{\specialrule{\lightrulewidth,gray7,dashed}{2.5pt}{2.5pt}}
\definecolor{lightb}{RGB}{235,245,255}

\newcommand{\mysection}[1]{\noindent\textbf{#1}}
\usepackage{algorithm}
\usepackage{algpseudocode}

\usepackage{listings}
\usepackage[frozencache,cachedir=.]{minted}
\usepackage{tcolorbox}
\usepackage{caption}
\usepackage{tabularx}

\newcommand\dscoderAbbrev{DS}

\newcommand{\modeld}{Magicoder-\dscoderAbbrev{}}
\newcommand{\modelx}{\mbox{Magicoder$\mathcal{S}$}}
\newcommand{\modelxd}{\modelx-\dscoderAbbrev}

\newcommand\gptthreefive{GPT-3.5}
\newcommand\gptthreefiveturb{\gptthreefive~Turbo}

\newcommand\gptfourturb{GPT-4~Turbo}

\newcommand\dscoderbase{DeepSeek-Coder-Base}
\newcommand\dscoderinst{DeepSeek-Coder-Instruct}

\newcommand\humaneval{HumanEval}

\newcommand\evalplus{EvalPlus}
\newcommand\mbpp{MBPP}

\title{Code Less, Align More: Efficient LLM Fine-tuning for Code Generation with Data Pruning}

\author{Yun-Da Tsai \\
  NVIDIA \\
  \texttt{yundat@nvidia.com} \\\And
  Mingjie Liu \\
  NVIDIA \\
  \texttt{mingjiel@nvidia.com}  \\\And
  Haoxing Ren \\
  NVIDIA \\
  \texttt{haoxingr@nvidia.com} \\}

\begin{document}
\maketitle
\begin{abstract}
Recent work targeting large language models (LLMs) for code generation demonstrated that increasing the amount of training data through synthetic code generation often leads to exceptional performance. 
In this paper we explore data pruning methods aimed at enhancing the efficiency of model training specifically for code LLMs. We present techniques that integrate various clustering and pruning metrics to selectively reduce training data without compromising the accuracy and functionality of the generated code. 
We observe significant redundancies in synthetic training data generation, where our experiments demonstrate that benchmark performance can be largely preserved by training on only 10\% of the data. Moreover, we observe consistent improvements in benchmark results through moderate pruning of the training data.
Our experiments show that these pruning strategies not only reduce the computational resources needed but also enhance the overall quality code generation. 
\end{abstract}

\section{Introduction}
\label{sec:intro}

The performance of large language models (LLMs) is heavily dependent on the size and quality of their training datasets, as highlighted by recent studies on scaling laws~\cite{achiam2023gpt, zhang2024scaling}. State-of-the-art code LLMs, such as CodeAlpaca~\cite{codealpaca}, WizardCoder~\cite{luo2024wizardcoder}, and MagicCoder~\cite{wei2023magicoder}, have achieved remarkable performance by significantly expanding their supervised fine-tuning datasets through synthetic code generation. Various synthetic code generation approaches have been developed, including the Self-Instruct technique~\cite{wang2022self}, Evol-Instruct~\cite{xu2023wizardlm}, and OSS-Instruct~\cite{wei2023magicoder}. However, such scaling approaches not only increase the training cost but also demands substantial computational resources, making it expensive and less accessible.

\nocite{tsai2023differential}

Achieving optimal performance in fine-tuned models for downstream tasks often relies on large, high-quality datasets. Recently, there has been a growing interest in more efficient fine-tuning methods for large language models (LLMs). One recent work introduces the Superficial Alignment Hypothesis~\cite{zhou2023lima}, which suggests that most knowledge in LLMs is acquired during pretraining, and only minimal instruction tuning data is required to align models with human preferences. Promising strategies to reduce computational demands include parameter-efficient fine-tuning (PEFT) methods, which reduce the number of parameters needed for training~\cite{fu2023effectiveness, hu2021lora}. Another research direction uses active learning to iteratively select data samples during training, thereby enhancing model learning~\cite{su2022selective, diao2023active}. These methods primarily aim to improve model accuracy through iterative processes, requiring multiple rounds of training and data selection.

\nocite{tsai2024toward}

Data selection and pruning methods have also been well-explored in literature, with evidence suggesting that careful pruning can sometimes even surpass the performance of using the full dataset~\cite{penedo2024fineweb,wang2023too}. 
\nocite{tsai2024handling}
Moreover, many of these methods are computationally intensive such as supervised metrics that involves multiple times of model training to keep track of loss and gradients~\cite{xia2024less,pruthi2020estimating} or heavy sampling method with Monte Carlo~\cite{schoch2023data}, limiting their scalability. Practical pruning methods that aims for large-scale data have been investigated in the contexts of LLM pretraining~\cite{das2023deft,penedo2024fineweb} and fine-tuning~\cite{chen2024alpagasus,schoch2023data} datasets, image datasets~\cite{moser2024study,meding2021trivial}, and vision-text training datasets~\cite{wang2023too}, and demonstrate success by applying clustering and by choosing proper indicator functions. 
\nocite{da2022fast}

Despite these advances, there remains a gap in efficient pruning strategies specifically tailored for coding datasets. Most large-scale code datasets are synthetically generated, resulting in many data samples with similar lexical appearances due to consistent formatting and style. Large-scale synthetic datasets commonly used for training code LLMs often suffer from significant redundancy and noise~\cite{wang2023too}. This redundancy arises from the impracticality of verifying the functional correctness of each program, leading to a substantial portion of instruction-code pairs being noisy. Therefore, enhancing data efficiency through careful selection and pruning of data samples is crucial for improving model performance without relying on excessively large datasets. 

In this work, we present a scalable and effective data pruning method to enhance code generation in large language models. Our approach clusters data samples based on problem instructions and their code solutions, applying dimensionality reduction to reduce computational load. We then select a representative subset from each cluster using various pruning metrics.
Experiments on large-scale datasets and evaluations on downstream coding tasks show that our method maintains or even improves model performance while significantly reducing training data. Our contributions and key findings are summarized as follows:

\begin{itemize}
    \item We are the first to study data pruning for large-scale synthetic code fine-tuning. We create an efficient and scalable pruning strategy based on unsupervised learning methods.
    \item We find large redundancies in synthetic generated code datasets, as training on just 10\% retains most benchmark performance, with slight degradation of 3.9\% on HumanEval and 1.5\% on MBPP compared with using all data.
    \item We observe consistent improvement by moderately pruning the dataset, leading to  improvements of up to 2.7\% on HumanEval and 3.5\% on MBPP compared with using all data.
    \item We perform detailed ablation studies, where results demonstrate the clustering algorithm to be critical, while pruning metrics to be less important. 
\end{itemize}

\section{Related Work}
\label{sec:related}

In this section, we review the advancements of large language models (LLMs) for code generation in Section~\ref{sec:related:codellm} and review prior work on instructional finetuning in Section~\ref{sec:related:finetune}. Finally, we discuss earlier research on data selection and pruning methods in Section~\ref{sec:related:pruning}.

\subsection{Large Language Models for Code Generation}
\label{sec:related:codellm}

Great advancements have been achieved in improving Large Language Models (LLMs) for code generation. Codealpaca~\cite{codealpaca} extends the capabilities of the LLaMA model~\cite{touvron2023llama} by incorporating 20,000 instruction-following data points generated through the Self-Instruct technique~\cite{wang2022self}, which aligns language models with self-generated instructions. CodeLlama~\cite{roziere2023code} further enhances this methodology by fine-tuning from LLaMA2~\cite{touvron2023llama2}, utilizing 14,000 instruction-following data points also generated via the Self-Instruct technique.
\nocite{tsai2021toward}

Wizardcoder~\cite{luo2024wizardcoder} utilizes the Evol-Instruct method~\cite{xu2023wizardlm} to evolve the Codealpaca dataset further. This technique iteratively evolves instruction-following data in both depth and breadth dimensions. On the other hand, Magicoder~\cite{wei2023magicoder} employs the OSS-Instruct technique to create instruction-following data from unlabeled open-source code snippets, constructing a dataset of 75,000 samples based on the StarCoder dataset~\cite{lozhkov2024starcoder}.

\subsection{Instructional Fine-tuning}
\label{sec:related:finetune}
Fine-tuning language models with instructional datasets has emerged as a powerful technique, offering notable improvements in model performance and alignment with human preferences and safety. By exploring a diverse array of instructional tasks, \cite{wei2021finetuned} demonstrated a significant enhancement in zero-shot performance on unseen tasks through fine-tuning. Building on this,  \cite{chung2024scaling} showed that scaling both the number of tasks and the model size can lead to substantial performance gains across different model architectures. \cite{peng2023instruction} further advanced this field by leveraging large language models (LLMs) to generate high-quality instruction-following data, resulting in improved zero-shot performance on new tasks. 

A recent study~\cite{zhou2023lima} introduces the Superficial Alignment Hypothesis, which posits that the bulk of knowledge in LLMs is acquired during pretraining. It further suggests that minimal fine-tuning data is sufficient to align these models with human preferences. The study demonstrates a noteworthy enhancement in LLM performance with just 1,000 high-quality instruction data points. Subsequently, a plethora of research endeavors have concentrated on refining dataset quality through diverse filtering methodologies for general instruction following~\cite{xu2023rethinking, chen2024alpagasus, liu2023makes}.
\nocite{tsai2024text}

\subsection{Data Pruning for Efficient Training}
\label{sec:related:pruning}




Various pruning methods have been explored for selecting more informative samples for model training, each tailored to different scenarios. 
Data clustering has been widely used as a highly effective technique for data pruning. TLDR~\cite{wang2023too} utilized KMeans clustering to group similar data points and uniformly sampled from each cluster. They employ Image-Text Matching (ITM) scores to identify suitable vision-text pairs, offering another perspective on sample selection. 
DEFT~\cite{das2023deft} utilizes unsupervised core-set selection for clustering-based data-efficient fine-tuning of LLMs. This approach significantly enhances data efficiency in fine-tuning for text-editing applications. 

Metrics like Hardness~\cite{sorscher2022beyond}, Instruction Following Difficulty (IFD)~\cite{li2023quantity} (Li et al., 2023), and SuperFiltering~\cite{li2024superfiltering} focus on identifying "hard" samples that are either difficult to learn or easy to forget, tracking each data sample throughout training.
In addition to these, sample influence metrics such as LESS~\cite{xia2024less} and TracIn~\cite{pruthi2020estimating} monitor model gradients and the impact of individual samples, albeit with significant computational overhead for large models and datasets.
Quality metrics from external oracles~\cite{chen2024alpagasus, liu2023makes}, leverage strong language models like ChatGPT for data selection. 
However, utilizing external oracles may not always be feasible due to cost constraints.

\begin{figure*}[ht]
\centering
  \centering
  \includegraphics[width=\linewidth]{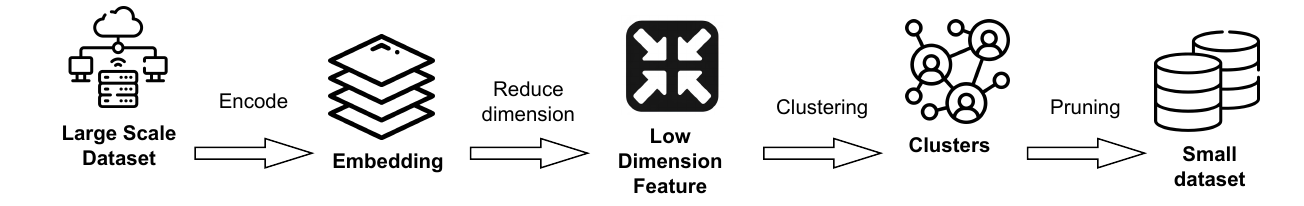}
  \caption{The overview of efficient data pruning for fine-tuning LLMs with large scale datasets. First, We reduce the encode instruction-following data into embedding and reduce the dimension of feature representation. Second, we apply clustering to identify and group up similar data samples. Finally, we applied pruning metrics to further reduce data size.
  }
  \label{fig:method}
\end{figure*}

\section{Methodology}
\label{sec:method}

Our goal is to select high-quality, representative data samples so that training on these subsets yields performance that is comparable to or better than training on the entire dataset. The overview of efficient data pruning for fine-tuning LLMs with large scale datasets is illustrate in Figure~\ref{fig:method}. First, we use an embedding model to project the instruction-code pairs into a vector representation. We further reduce the dimension of feature representation to reduce computation complexity of the following steps. We then apply clustering to identify and group up similar data samples. Finally, we applied pruning metrics to further reduce data size. The detail pseudo code is in Algorithm~\ref{alg:pruning}.

When dealing with coding datasets, two primary selection directions can be considered: syntactical and semantic. Selecting programs that are syntactically different but semantically equivalent, or vice versa, can be inefficient.
Our design will focus on identifying syntactical differences. Detecting semantic differences between programs typically requires fuzzing techniques~\cite{chen2018systematic}, which involve creating larger test samples and executing programs to group them based on behavior. This approach contradicts our objective of reducing computational costs. Therefore, our method emphasizes syntactical analysis to achieve efficient and effective data selection.

\begin{algorithm}[ht]
\caption{Data Pruning Algorithm}\label{alg:pruning}
\begin{algorithmic}[1]
\State Initialize $Embbedding$, Compression $Ratio$
\State Initialize $selected \gets []$
\State $X \gets$ PCA($Embedding$)
\State $Cluster \gets$ ClusterAlgo($X$)

\For{each $idx, items$ in $Cluster$}
    \State $score \gets$ PruningMetrics($item$)
    \State $remain \gets$ Random($items$, prob=$score$)
    \State Update $Cluster[ids] \gets remain$
    \State Append $selected \gets remain$
\EndFor

\State \textbf{Output:} $selected$

\end{algorithmic}
\end{algorithm}

\subsection{Dimension Reduction}
We convert each instruction-code pair into vector representation using a embedding model from raw text to enhance the efficiency of clustering and computation of pruning metrics~\cite{naik2024limitations}. Recent research indicates that distances based on LLM embeddings effectively capture syntactic differences. To address the computational complexity, we employ Principle Component Analysis (PCA) \cite{mackiewicz1993principal} to reduce the dimensionality of the vector representations, as representations extracted from LLMs often exceed a thousand dimensions. Moreover, this approach prevents the subsequent utilization of several pruning metrics, which involve kernel methods, from being hindered in high-dimensional spaces by the curse of dimensionality.
\nocite{tsai2023rtlfixer}

\subsection{Clustering}
Clustering is a critical step in our methodology to group similar instruction-code pairs, which facilitates the selection of diverse and representative samples.  Before clustering, we normalize the vector representations to ensure that each feature contributes equally to the distance calculations. From each cluster, we then sample instruction-code pairs to create a subset that is representative of the entire dataset. The sampling strategy is further decided by different pruning metrics. 

\subsubsection{KMeans}
The KMeans algorithm~\cite{kanungo2002efficient} partitions data into \(k\) clusters.
By minimizing the within-cluster sum-of-squares, KMeans ensures that each cluster is as compact as possible.
The main advantage of KMeans is its scalability and efficiency in handling large datasets.

\subsubsection{Agglomerative Clustering}
Agglomerative Clustering~\cite{mullner2011modern} builds nested clusters with linkage criteria. This method is advantageous since it does not require the number of clusters to be specified a priori. This flexibility allows for a more nuanced selection of representative samples, which is beneficial for maintaining the quality of the dataset.

\subsubsection{HDBSCAN}
Hierarchical Density-Based Spatial Clustering of Applications with Noise (HDBSCAN)~\cite{rahman2016hdbscan} performs clustering based on the concept of core samples, which are samples located in high-density areas measured by a distance metric. This approach aligns well with our design hypothesis to find the most syntactically representative data samples. Notably, HDBSCAN removes noisy samples not clustered into core samples as outliers.

\begin{table*}[ht]
    \centering
         \begin{tabular}{
        @{}lccccc
    }
        \toprule
        \multicolumn{1}{l}{Model}  & \multicolumn{1}{c}{Training} & \multicolumn{2}{c}{Benchmark} & \multicolumn{2}{c}{Improvement Over Base}\\
        \cmidrule(lr){3-4} \cmidrule(lr){5-6}
        & Tokens & \humaneval~(+) & \mbpp~(+) & \humaneval~(+) & \mbpp~(+) \\
        \midrule
        \gptthreefiveturb  & - & 72.6~~(65.9) & 81.7~~(69.4) & - & - \\
        \gptfourturb & - & \textbf{85.4}~~(\textbf{81.7}) & \textbf{83.0}~~(\textbf{70.7}) & - & - \\
        \midrule
        
        \dscoderbase & - & 47.6~~(39.6)  & 70.2~~(56.6) & - & - \\
        \dscoderinst  & 2B & 73.8~~(70.1)  & 72.7~~(63.4) & 26.2~~(30.5) & 2.5~~(6.8) \\
        \modeld   & 90M & 66.5~~(60.4) & 75.4~~(61.9) & 18.9~~(20.8) & 5.2~~(5.3) \\
        \modelxd   & 240M & \textbf{76.8}~~(\textbf{70.7}) & \textbf{75.7}~~(\textbf{64.4}) & \textbf{29.2}~~(\textbf{31.1}) & \textbf{5.5}~~(\textbf{7.8}) \\
        \midrule
        Ours (full data) & 234M & 74.3~~(70.8) & 74.5~~(62.3) & 26.7~~(31.2) & 4.3~~(5.7) \\
        Ours (90\%) & 192M & \textbf{77.0}~~(\textbf{71.6}) & 76.9~~(\textbf{64.0}) & \textbf{29.4}~~(\textbf{32.0}) & 6.7~~(\textbf{7.4}) \\
        Ours (50\%) & 106M & 71.0~~(64.0) & \textbf{78.0}~~(\textbf{64.0}) & 23.4~~(24.4) & \textbf{7.8}~~(\textbf{7.4}) \\
        Ours (10\%) & 21M & 70.4~~(65.0) & 73.0~~(60.2) & 22.8~~(25.4) & 2.8~~(3.6) \\
        Ours (1\%) & 2M & 64.6~~(58.0) & 74.3~~(61.9) & 17.0~~(18.4) & 4.1~~(5.3) \\
        \bottomrule
    \end{tabular}
    
    \caption{$pass@1$ (\%) results of different LLMs on \humaneval{}~(+) and \mbpp{}~(+) with greedy decoding. We directly use results from prior work~\cite{deepseek-coder, wei2023magicoder}. All our results are reported using the HDBSCAN clustering algorithm with the diversity pruning metric (HDBSCAN-diversity). To account for the randomness of clustering and training, we report the averaged results from three runs evaluated with \evalplus~\cite{evalplus}.
     }
    \label{tab:python-text2code}
\end{table*}

\subsection{Pruning Metrics}
The criteria of choosing pruning metrics continually aligns with the idea of detecting syntactic difference and find most representative samples. We explain the pruning metrics explored in our experiments in the following sections.

\subsubsection{Diversity Metric}
We use a distance-based metric that simply evaluates the diversity score of a single instance shown as follow,
\begin{equation}
    d_i = \min_{\mathbf{x} \in \mathcal{K} \setminus \{\mathbf{x}_i\}} \text{dist}(\mathbf{x}_i, \mathbf{x}),
\end{equation}
where $x_i$ is the vector representation, \textit{dist} is a distance function, $K$ represents selected query set within the dataset cluster, and $d_i$ is the diversity score of a sample $x_i$. We use the dot product of the embeddings as the distance function as our embeddings are normalized prior to pruning.

\subsubsection{Density Metric}
We applied kernel density estimation (KDE) to measure the density of samples in the feature space. KDE estimates the probability density function of a random variable. The density score for a sample $\mathbf{x}_i$ is given by,
\begin{equation}
    \rho(\mathbf{x}_i) = \frac{1}{n h^d} \sum_{j=1}^{n} K\left(\frac{\mathbf{x}_i - \mathbf{x}_j}{h}\right),
\end{equation}
where $K$ is the kernel function, $h$ is the bandwidth parameter, $d$ is the dimension of the feature space, and $n$ is the total number of samples. The kernel function $K$ (typically a Gaussian) measures the influence of nearby points on the density estimate.
A high density score indicates that a sample is located in a region with many similar instances, suggesting it is less critical for maintaining diversity.

\subsubsection{Random}
The simplest baseline is random selection, where we randomly sample data from the selected cluster or entire training dataset (without clustering) for instruction tuning. 

\begin{figure*}[ht]
\centering

\begin{subfigure}[b]{0.48\linewidth}
  \centering
  \includegraphics[width=\linewidth]{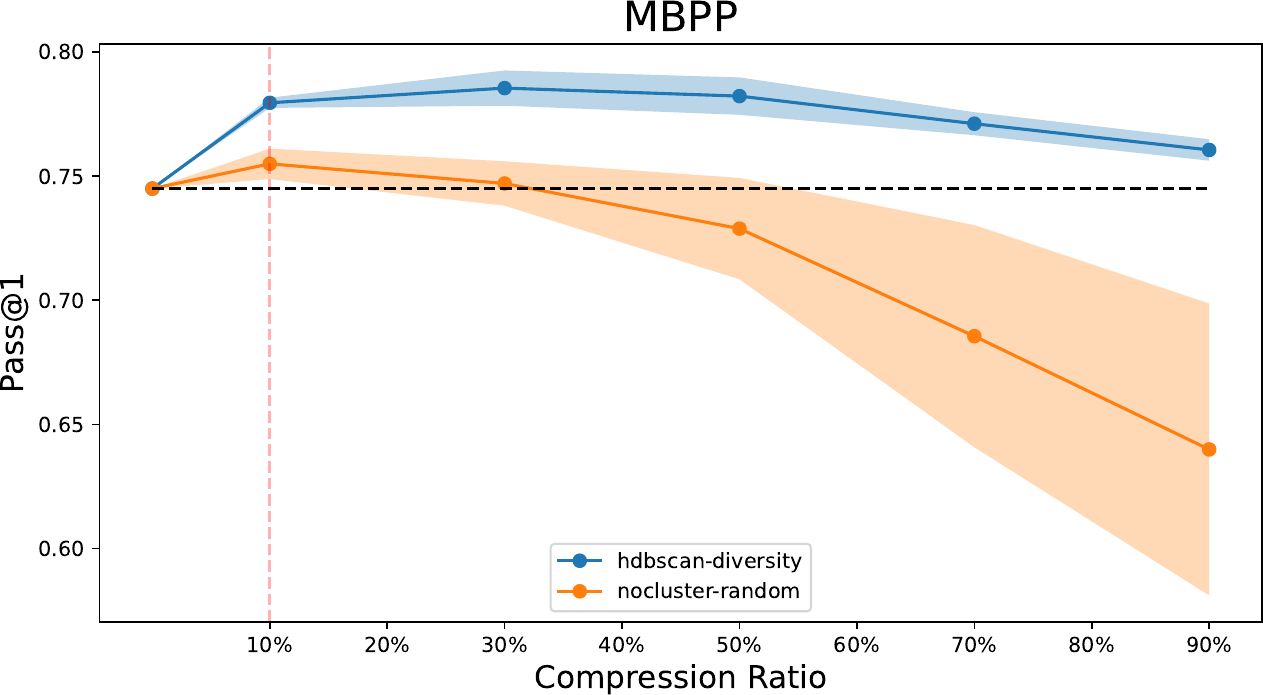}
  \label{fig:exp:main-mbpp}
\end{subfigure}
\hfill
\begin{subfigure}[b]{0.48\linewidth}
  \centering
  \includegraphics[width=\linewidth]{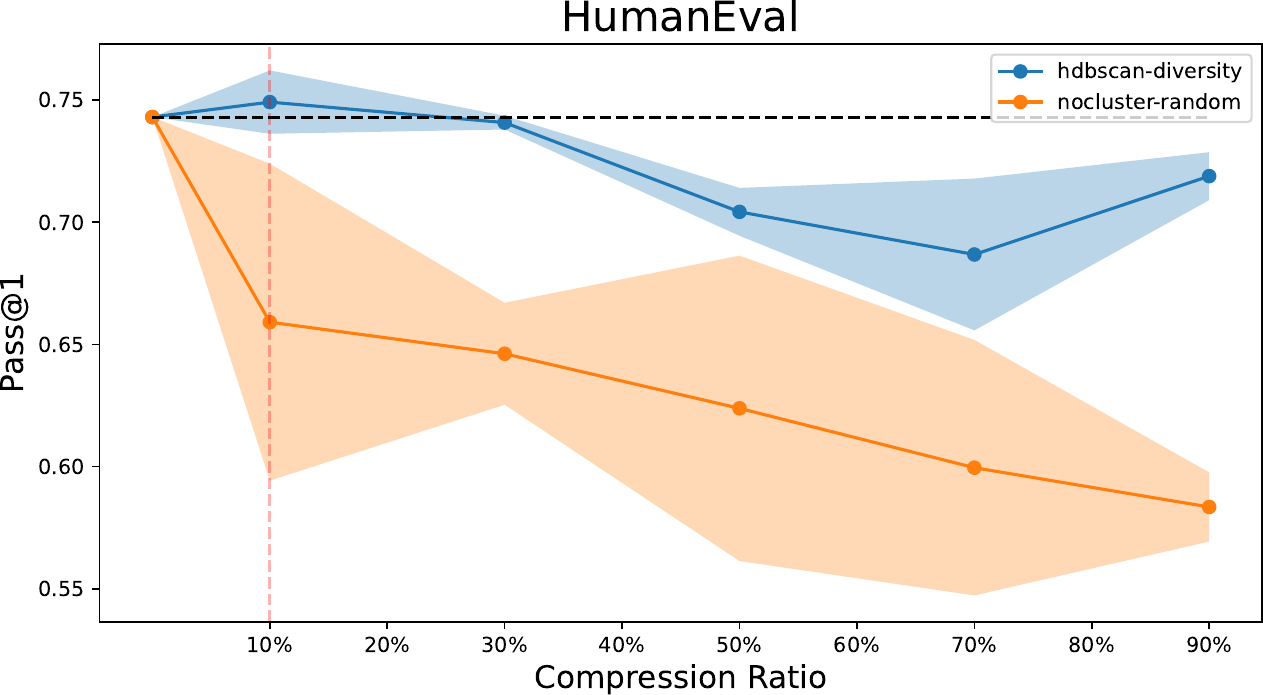}
  \label{fig:exp:main-humaneval}
\end{subfigure}



\caption{Performance comparison of HDBSCAN-diversity and nocluster-random methods across different benchmarks. Our strategy outperform the baseline across different datasets with a large margin. We also maintain better or equivalent performance compare to full dataset even at the size of 10\% on MBPP. The $pass@1$ metric is plotted against varying compression ratios, demonstrating the robustness and effectiveness. HumanEval presents larger variance across experiments possibly due to less problems entries.}
\label{fig:exp:main-results}
\end{figure*}

\section{Experiments}
\label{sec:exp}

In this section, we first present the experimental setup in Section~\ref{sec:exp:setup}, followed by our primary findings in Section~\ref{sec:exp:main}. Here, we highlight the performance improvements of our pruning methods compared to full dataset training across four datasets: MBPP(+), and HumanEval(+). We also compare the $pass@1$ scores with baseline methods at various compression ratios.

\subsection{Setup}
\label{sec:exp:setup}
We employed DeepSeek-Coder-Base 6.7B~\cite{deepseek-coder} as the base model due to its superior performance among open-source models. We used PCA~\cite{mackiewicz1993principal} algorithm in all experiments and reduce the dimension to 10. To account for randomness in clustering algorithm and training, we repeat each experiment 3 times and report the average and standard deviation.

\subsection{Training}
\label{sec:exp:train}

\mysection{Datasets}
\label{sec:exp:setup:training}
In our experiment, we adopt two synthetic code dataset as training data: Magicoder-OSS-Instruct-75K~\footnote{\url{https://huggingface.co/datasets/ise-uiuc/Magicoder-OSS-Instruct-75K}} (MIT License) and Magicoder-Evol-Instruct-110K~\footnote{\url{https://huggingface.co/datasets/ise-uiuc/Magicoder-Evol-Instruct-110K}} (Apache-2.0 License).
Together we have a combined 185k entries in total as our target large scale dataset.

We fine-tune the base model by combining and shuffling the two training dataset.
This is different as in the original Magicoder~\cite{wei2023magicoder} implementation, where they first fine-tune the base models for 2 epochs on OSS-Instruct data and continue training for 2 more epochs on Evol-Instruct data. 
We note that despite such difference in our implementation details, our full dataset performance closely matches the \modelxd  \ results.

\begin{figure*}[htbp]
\centering
\begin{subfigure}[b]{0.48\linewidth}
  \centering
  \includegraphics[width=\linewidth]{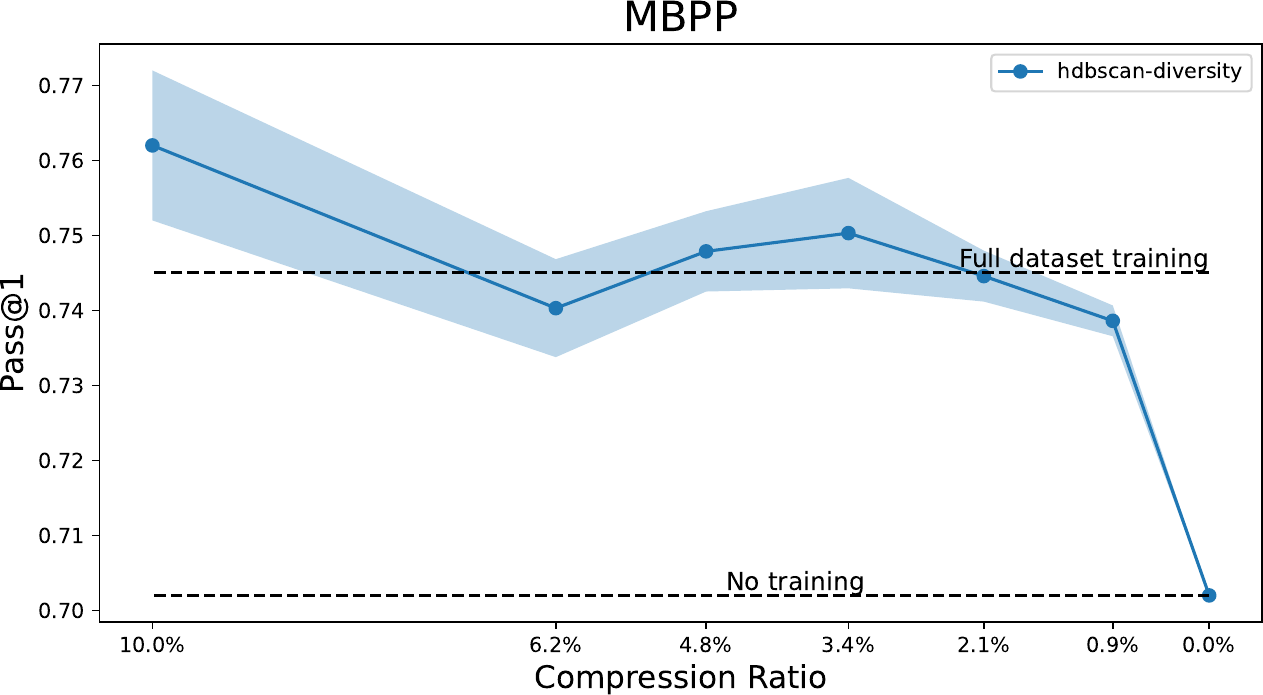}
\end{subfigure}
\hfill
\begin{subfigure}[b]{0.48\linewidth}
  \centering
  \includegraphics[width=\linewidth]{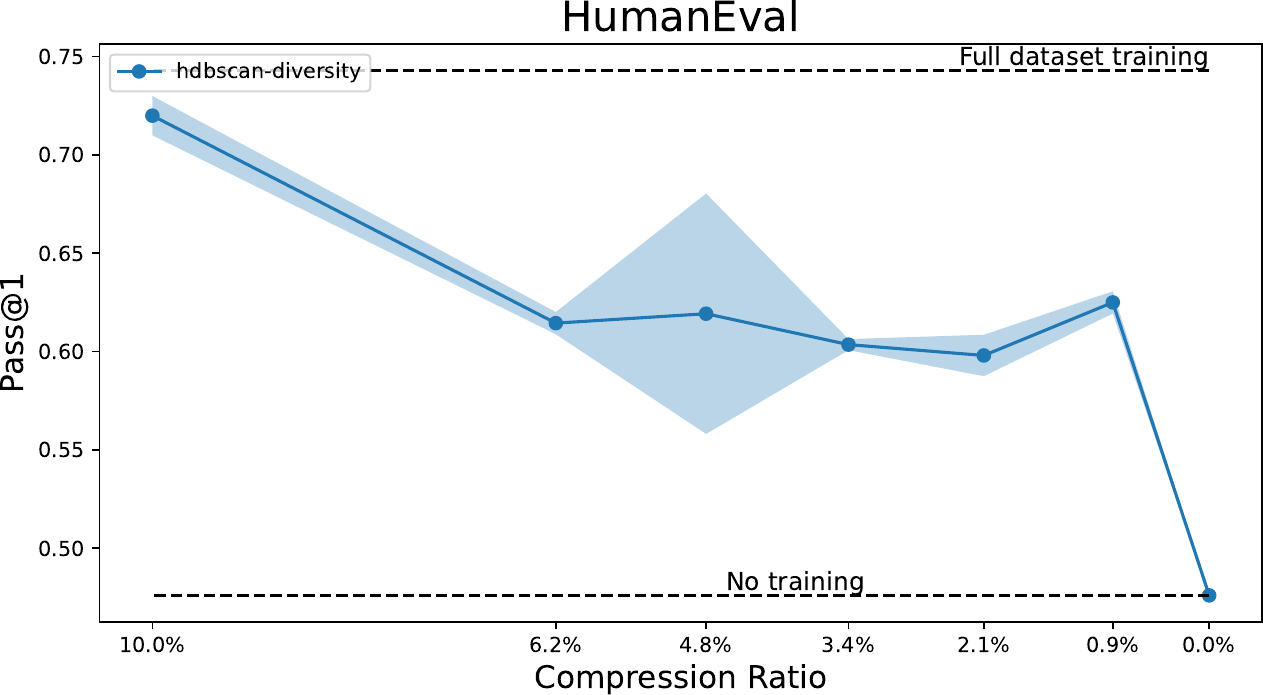}
\end{subfigure}
\caption{Comparison of performance under extreme data pruning conditions on the MBPP and HumanEval benchmarks. The $pass@1$ score on MBPP shows that even with just 1\% of the data, our method achieves nearly equivalent performance to the full dataset, with a 4.1\% improvement over the base model. On the HumanEval benchmark, while the performance with 1\% of the data degrades compared to the full dataset training, it still achieves an 17.0\% improvement over the base model.}
\label{fig:extreme-side-by-side}
\end{figure*}

\mysection{Training}
Training is conducted with 16 NVIDIA A100-80GB GPUs through the Distributed Data Parallel (DDP) module from PyTorch. We set the learning rate at 5e-5 with 15 warmup steps and a linear learning rate scheduler. We use Adam~\cite{kingma2014adam} as our optimizer with full parameter updates and truncate sequence length longer than 4096 tokens.
We use a batch size of 512 samples~\cite{wei2023magicoder} when the dataset size exceeds $\geq 10\%$ of the original size, and a batch size of 32~\cite{zhou2023lima} for heavily pruned small-scaled data experiments in Figure~\ref{fig:extreme-side-by-side}.
We fine-tune for 2 epochs regardless of the dataset size.

\subsection{Evaluation}
\label{sec:exp:eval}

\mysection{Datasets}
HumanEval~\cite{chen2021evaluating} and MBPP~\cite{austin2021program} are two of the most widely used benchmarks for code generation. The two datasets contains 164 and 1401 problems respectively. Each task in these benchmarks includes a task description (e.g., docstring) as the prompt, where LLMs generate corresponding code whose correctness is checked by a handful of test cases. Because tests in these benchmarks can be insufficient, for more rigorous evaluation, we use HumanEval+ and MBPP+, both powered by EvalPlus~\cite{evalplus} to obtain 80× and 35× more tests, respectively.

\mysection{Metric}
Following prior work~\cite{chen2021evaluating,evalplus}, for each experiment we use the unbiased pass@k estimator shown as follow and mainly focus on comparing $pass@1$ metric:
\begin{equation}
    pass@k := \mathbb{E}_{\text{Problems}} \left[ 1 - \frac{\binom{n-c}{k}}{\binom{n}{k}} \right].
\end{equation}
 
\mysection{Inference}
We employ the EvalPlus~\cite{evalplus} inference script with sanitation postprocessing. We adopted the vLLM~\cite{kwon2023efficient} framework and use greedy decoding for every code generation. The inference engine is setup with bf16 dtype, tensor parallel size of 2 and a maximum length of 4096.

\subsection{Implementation Details}
\label{sec:exp:detail}
In our experiment, the PCA reduction is fitted on the benchmark dataset and then apply the projection to the instruction data. We used the OpenAI \textit{text-embedding-ada-002} embedding model to encode data. All the clustering and kernel density estimation parameters are as default in sklearn~\cite{scikit-learn}. For algorithms that requires choosing an optimal number of clusters (such as KMeans) is crucial, we utilize the Elbow method~\cite{roy1953heuristic} to find the point where adding more clusters does not significantly improve the variance explained. For pruning metrics, we applied the Scott's Rule~\cite{scott2010scott}, a normal-reference rule for deciding the Gaussian kernel bandwidth, for kernel density estimation and random select 10\% of the dataset as query set ($K$) for diversity metric.

\subsection{Main Results}
\label{sec:exp:main}

Table~\ref{tab:python-text2code} presents the $pass@1$ results of different leading code LLMs on the HumanEval and MBPP benchmarks, computed with greedy decoding. All our results are reported using the HDBSCAN clustering algorithm with the diversity pruning metric (HDBSCAN-diversity). To account for the randomness of clustering and training, we report the averaged results from three runs. 
Notably, slight pruning of the training data could yield a performance improvement of up to 2.7\% on HumanEval and 3.5\% on MBPP compared to training with the full dataset. We further show that benchmark accuracy can be largely retained with 10\% of the dataset, with slight degradation of 3.9\% on HumanEval and 1.5\% on MBPP compared with using the full training data. Even with just 1\% of the data ($\sim$ 700 samples), our method maintains competitive performance and achieves large improvements over the base model, underscoring the efficiency of our pruning strategy.

Figure~\ref{fig:exp:main-results} illustrates the detail of our pruning methods across four datasets: MBPP, MBPP+, HumanEval, and HumanEval+. Each subplot compares the $pass@1$ scores of the HDBSCAN-diversity method with the nocluster-random baseline at various compression ratios. HDBSCAN-diversity method consistently outperforms the nocluster-random baseline. The performance typically improves with slight compression, peaking around 10-20\%, and then gradually declines. This trend highlights the robustness of the HDBSCAN-diversity method, maintaining higher $pass@1$ scores than full dataset even at 90\% compression. 

We further examine how our data pruning method performs when pushed to the extreme, aiming to achieve the smallest possible dataset size on the MBPP benchmark. 
The results are presented in Figure~\ref{fig:extreme-side-by-side}. Remarkably, we found that even with just 1\% of the data, our method achieves a 4.1\% improvement over the base model, which is nearly equivalent to training on the full dataset. This demonstrates the robustness of our pruning method, highlighting its ability to maintain high performance with minimal data, thus significantly reducing the computational resources required. 

Overall, these results demonstrate the effectiveness of data pruning strategy in preserving critical data features and maintaining model performance under significant data reduction, making it a superior choice for coding dataset pruning.

\section{Ablation Studies}

Our research includes four ablation studies designed to evaluate the impact of (1) clustering algorithms (2) pruning metrics (3) dimension reduction (4) input for vector representation on the effectiveness of data pruning. In the studies, we will mainly focus on the MBPP benchmark since it provides more stable and consistent results.

\subsection{Compare Clustering Algorithm}
\begin{figure}[t!]
\centering
  \centering
  \includegraphics[width=\linewidth]{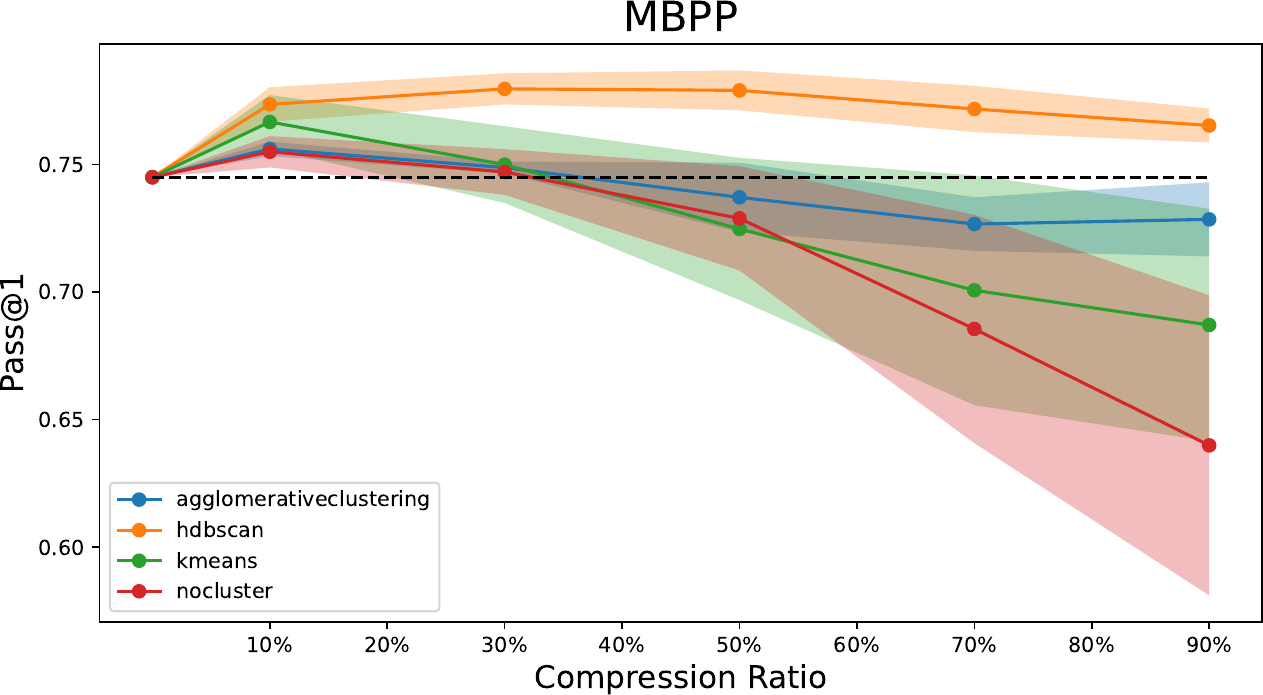}
  \caption{$pass@1$ on the MBPP benchmark comparing across different clustering algorithms and varied compression ratios of the training dataset. HDBSCAN demonstrate strong robustness in maintaining higher $pass@1$ scores compared to full dataset at the compression ratio of 90\%.}
  \label{fig:ablation:cluster}
\end{figure}

In Figure~\ref{fig:ablation:cluster}, we present the results of applying different clustering algorithms without additional pruning metrics. The algorithms evaluated include Agglomerative Clustering, HDBSCAN, KMeans, and a baseline with no clustering (nocluster).

The results demonstrate that clustering algorithms generally improve performance compared to the nocluster baseline, particularly at higher compression ratios. HDBSCAN consistently maintains higher $pass@1$ scores, showcasing its robustness in preserving critical data features. KMeans and Agglomerative Clustering also perform well, though with higher variability. These findings highlight the importance of clustering algorithms in enhancing data efficiency for coding datasets.

\begin{figure}[htbp]
  \centering
  \includegraphics[width=\linewidth]{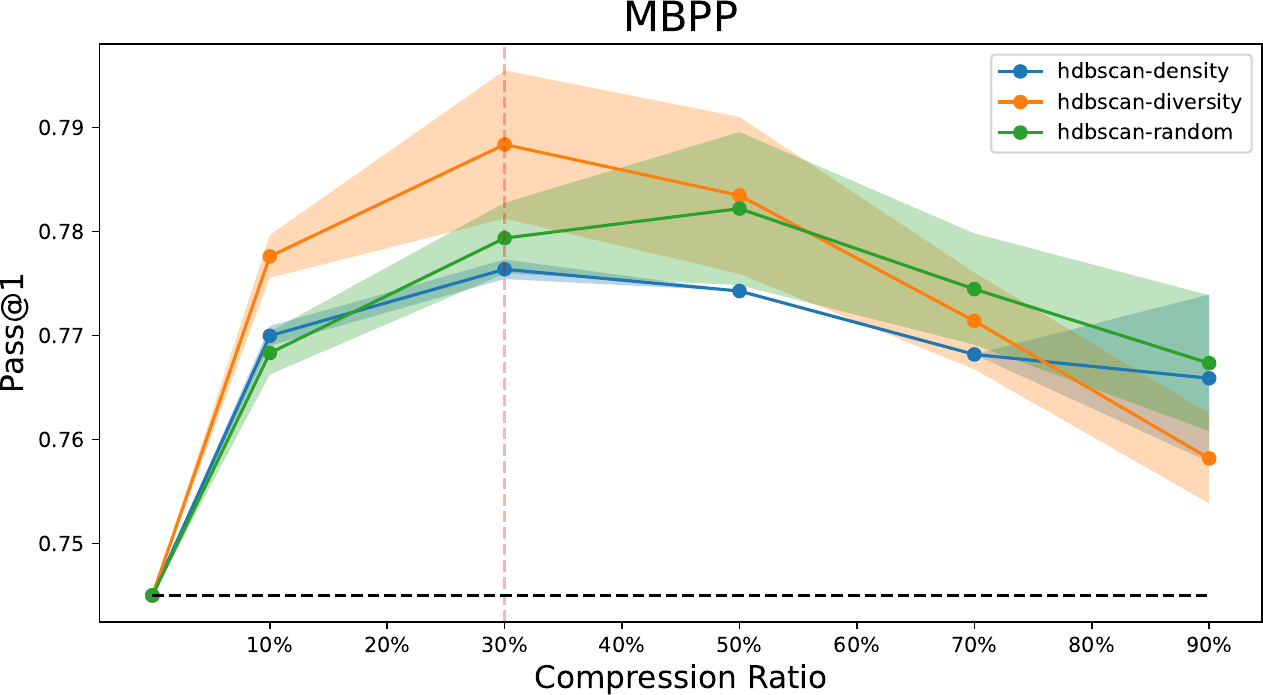}
\caption{Comparison of different pruning metrics using HDBSCAN clustering algorithms. Diversity metric has marginal advantage but its benefit may be limited and dependent on the clustering algorithm.}
\label{fig:ablation:prune}
\end{figure}

\subsection{Compare Pruning Metrics}
We examine the impact of different pruning metrics on model performance. Using HDBSCAN clustering algorithm, we assess how these metrics influence performance as the data size decreases, as illustrated in Figure~\ref{fig:ablation:prune}.
The results indicate that the effectiveness of pruning metrics varies across different compression ratio. While Diversity metrics show slight improvements over other metrics, the margin of improvement is not substantial and only works between 10-40\% compression ratio. This suggests that while more sophisticated pruning metrics can offer some benefits, their impact may be limited and also dependent on the clustering algorithm used.

\subsection{Effect of PCA}
In Table~\ref{tab:ablation:pca}, we evaluate the impact of applying Principal Component Analysis (PCA) on the performance of the KMeans clustering algorithm and Density metric at the compression ratio of 50\%.
The findings indicate that applying PCA generally degrades performance in terms of $pass@1$ scores for less than 0.6\% on MBPP, and moderate negative impact of 4.3\% on HumanEval. We hypothesize that the observed impact might be due to the imbalance between the MBPP and HumanEval datasets used for PCA training. Since the HumanEval dataset is significantly smaller than the MBPP dataset, it results in suboptimal extraction of principal components for HumanEval-like data. 

Nonetheless, reducing the dimension from 1536 to 10 leads to $\sim$12x speed up for KMeans. HDBSCAN clustering without PCA does not complete within 4 hours, thus we do not report its numbers.

\begin{table}[t!]
\centering
\begin{tabular}{lcc}
\toprule
 & No PCA & PCA \\
\midrule
Dimension & 1536 & 10 \\
Runtime & 1307 sec & 183 sec \\
\mbpp~(+) & 74.4~(63.3) & 73.8~(62.4) \\
\humaneval~(+) & 71.8~(65.0) & 67.5~(62.5) \\
\bottomrule
\end{tabular}
\caption{Comparison of $pass@1$ scores, dimension, and data pruning runtime (excludes embedding and training) at 50\% compression ratio for KMeans clustering with and without PCA (averaged over 3 runs).}
\label{tab:ablation:pca}
\end{table}

\subsection{Embeddings for Instruction or Code}
In Table~\ref{tab:ablation:feature}, we investigate the influence of various inputs on the embedding model. Specifically, we examine the effects of using only the instruction, only the code solution, or both as inputs for generating embeddings.
Our findings indicate that combining both instructions and code as embedding inputs yields better performance compared to using either one alone. There are no significant differences in the results when using only instructions or only code. This suggests that even though instructions and code samples often correspond closely, it is crucial to maintain diversity and select informative samples from both during data pruning.



\begin{table}[h!]
\centering
\begin{tabular}{lcc}
\toprule
Feature Type & \mbpp~(+) & \humaneval~(+) \\
\midrule
Both & 76.3 (62.5) & 73.1 (69.6) \\
Instruction & 74.0 (63.7) & 69.1 (63.6) \\
Code & 74.1 (62.7) & 69.2 (63.3) \\
\bottomrule
\end{tabular}
\caption{$pass@1$ scores for different embedding inputs with 50\% compression ratio using KMeans clustering. Using both instruction and code brings slight benefits.}
\label{tab:ablation:feature}
\end{table}

\section{Conclusion}
\label{sec:conclusion}
This study presents an efficient data pruning strategy designed to improve the efficiency of fine-tuning large language models on coding datasets. Our results demonstrate that advanced clustering and pruning techniques can significantly improve data efficiency in LLMs, reducing computational costs while maintaining performance. Future work could focus on enhancing data quality by generating more informative data from clusters with low pruning metrics. We hope our findings provide valuable insights for developing more effective and scalable strategies in training code-focused LLMs, further enhancing synthetic data generation and the efficiency of human annotations.

\clearpage

\section*{Limitations}
One of the key limitations of our study is the inherent randomness from the clustering algorithms and training framework. Due to computational constraints, we only performed three runs and averaged the results for each of our experiments. While this approach provides a general indication of performance, it may not fully capture the variability and could lead to less accurate conclusions. More extensive experimentation with a larger number of runs would be necessary to achieve a higher degree of confidence in the results.

Throughout our experiments, we closely follow the hyperparameters described in~\cite{wei2023magicoder}, using a batch size of 512 samples and training for 2 epochs. However, such a high batch size results in only a few gradient updates when training on smaller datasets. Therefore, we switch to a lower batch size of 32, as recommended in~\cite{zhou2023lima}, when our pruned dataset is less than 10\% of the original size. 
We acknowledge that different hyperparameter settings could affect training outcomes and defer the determination of optimal hyperparameter settings for various training data sizes as future work.

\section*{Potential Risks}
This study focus exclusively on English prompts for Python code generation, thus prompts in other languages might not produce accurate or functional code. Additionally, the lack of safety alignment means there is a risk of generating malicious code or harmful language, which could lead to security vulnerabilities or unintended consequences. Code generation using LLMs carries inherent risks, such as producing incorrect or suboptimal code, failing to adhere to best practices, or introducing security flaws. Furthermore, LLMs may inadvertently propagate biases present in their training data, leading to biased outcomes in the generated code.

\section*{Use of AI Assistants}
ChatGPT was utilized to refine paper writing and generate code templates for drawing figures. The authors took careful attention to ensure that AI-generated contents are accurate and align with the authors intentions.



\bibliographystyle{unsrtnat}
\bibliography{custom}

\clearpage
\onecolumn

\appendix
\section{Code Samples from Data Pruning}
\label{sec:appendix}
We show examples from our data pruning. The selected data samples adheres closely to the downstream coding tasks, from English problem description to code generation. We also provide an example of removed data from our pruning strategy.

\begin{figure*}[h!]

\begin{center}

\begin{tcolorbox}[width=1.0\linewidth, halign=left, colframe=black, colback=white, boxsep=0.01mm, arc=1.5mm, left=2mm, right=2mm, boxrule=0.5pt]\footnotesize


\textbf{Input}\\

You are given a Python function that takes in a list of integers and a target integer. The function is intended to find two distinct elements in the list that sum up to the target integer. However, the function is incomplete and needs to be filled in with the appropriate code to achieve this functionality.

Complete the Python function `find\_two\_sum` to find and return the indices of two distinct elements in the input list that sum up to the target integer. If no such pair exists, return an empty list.
The function signature is:

\begin{minted}[breaklines]{python}
def find_two_sum(nums: List[int], target: int) -> List[int]:
\end{minted}

For example, given the input `nums = [2, 7, 11, 15]' and `target = 9', the function should return `[0, 1]` since `nums[0] + nums[1] = 2 + 7 = 9'.

\vspace{0.3in}
\textbf{Output}
\begin{minted}[breaklines]{python}
from typing import List

def find_two_sum(nums: List[int], target: int) -> List[int]:
    num_indices = {}
    for i, num in enumerate(nums):
        complement = target - num
        if complement in num_indices:
            return [num_indices[complement], i]
        num_indices[num] = i
    return []
\end{minted}
\end{tcolorbox}

\end{center}
\caption{Example of a selected data sample.}

\end{figure*}

\begin{figure*}[ht]

\begin{center}

\begin{tcolorbox}[width=1.0\linewidth, halign=left, colframe=black, colback=white, boxsep=0.01mm, arc=1.5mm, left=2mm, right=2mm, boxrule=0.5pt]\footnotesize


\textbf{Input}\\
You are given a list of integers representing the daily stock prices of a company. You need to write a function to calculate the maximum profit that can be obtained by buying and selling the stock at most once. If it is not possible to make any profit, return 0.

Write a function `maxProfit' that takes in a list of integers `prices` as input and returns the maximum profit that can be obtained.\\

Function signature:
\vspace{0.1in}

\begin{minted}[breaklines]{python}
def maxProfit(prices: List[int]) -> int:
\end{minted}

Example:
\begin{minted}[breaklines]{text}
Input: prices = [7, 1, 5, 3, 6, 4]
Output: 5
Explanation: Buy on day 2 (price = 1) and sell on day 5 (price = 6), profit = 6 - 1 = 5.

Input: prices = [7, 6, 4, 3, 1]
Output: 0
Explanation: In this case, no transaction is done, i.e., max profit = 0.
\end{minted}

Assumptions:\\
- The input list `prices' will have at least one element.\\
- The stock prices are given in chronological order.\\

\vspace{0.3in}
\textbf{Output}
\begin{minted}[breaklines]{python}
from typing import List

def maxProfit(prices: List[int]) -> int:
    if not prices:
        return 0
    
    min_price = prices[0]
    max_profit = 0
    
    for price in prices:
        if price < min_price:
            min_price = price
        else:
            max_profit = max(max_profit, price - min_price)
    
    return max_profit
\end{minted}

The `maxProfit' function iterates through the list of prices, keeping track of the minimum price encountered so far and the maximum profit that can be obtained. It updates these values as it iterates through the list and returns the maximum profit at the end.
\end{tcolorbox}
\end{center}
\caption{Another example of a selected data sample.}
\end{figure*}

\begin{figure*}[ht]

\begin{center}

\begin{tcolorbox}[width=1.0\linewidth, halign=left, colframe=black, colback=white, boxsep=0.01mm, arc=1.5mm, left=2mm, right=2mm, boxrule=0.5pt]\footnotesize


\textbf{Input}\\
\begin{minted}[breaklines]{text}
Background:
You are a Product Manager, named Alice, your goal is Efficiently create a successful product, and the constraint is . 
Question:

# Context
## Original Requirements
[BOSS: Write a podcast search engine which search for all index podcast. Each podcast belongs to a channel and each podcast contain title, description, published, audio link, transcript text]

## Search Information
### Search Results
### Search Summary

## mermaid quadrantChart code syntax example. DONT USE QUOTO IN CODE DUE TO INVALID SYNTAX. Replace the <Campain X> with REAL COMPETITOR NAME

quadrantChart
    title Reach and engagement of campaigns
    x-axis Low Reach --> High Reach
    y-axis Low Engagement --> High Engagement
...
\end{minted}

\vspace{0.1in}
\textbf{Output}\\
\begin{minted}[breaklines]{text}
## Original Requirements
We are tasked with creating a podcast search engine that indexes all podcasts. Each podcast is associated with a channel and contains a title, description, publication date, audio link, and transcript text.

## Product Goals
```python
[
    "Create a user-friendly podcast search engine that indexes all podcasts",
    "Ensure each podcast contains necessary information: title, description, publication date, audio link, and transcript text",
    "Develop a system that efficiently updates and maintains the podcast index"
]
```

## User Stories
```python
[
    "As a user, I want to search for podcasts by title or description so that I can find the content I'm interested in",
    "As a user, I want to see all the necessary information about a podcast (title, description, publication date, audio link, transcript text) so that I can decide whether to listen to it",
    "As a user, I want the search engine to suggest podcasts based on my search history so that I can discover new content",
    "As a user, I want the podcast search engine to be updated regularly so that I can access the latest podcasts",
    "As a user, I want to filter search results by publication date, channel, or transcript text so that I can narrow down my search"
]
```

## Competitive Analysis
...
\end{minted}

\end{tcolorbox}

\end{center}
\caption{Example of a removed data sample (outlier).}

\end{figure*}


\end{document}